\documentclass[%
 reprint,
superscriptaddress,   
 amsmath,amssymb,
 aps,
prl,
]{revtex4-2}

\usepackage{graphicx}
\usepackage{dcolumn}
\usepackage{bm}

\UseRawInputEncoding
\begin{document}

\preprint{APS/123-QED}

\title{Stand-off magnetometry with directional emission from sodium vapors}

\author{Rui Zhang}
\affiliation{State Key Laboratory of Advanced Optical Communication Systems and Networks, Department of Electronics, and Center for Quantum Information Technology, Peking University, Beijing 100871, China}
\affiliation{Beijing Academy of Quantum Information Sciences, Beijing 100193, China}
\affiliation{Johannes Gutenberg-Universit{\"a}t Mainz, 55128 Mainz, Germany}
 \affiliation{Helmholtz-Institut Mainz, GSI Helmholtzzentrum f{\"u}r Schwerionenforschung, 55128 Mainz, Germany}

\author{Emmanuel Klinger}
\affiliation{Johannes Gutenberg-Universit{\"a}t Mainz, 55128 Mainz, Germany}
 \affiliation{Helmholtz-Institut Mainz, GSI Helmholtzzentrum f{\"u}r Schwerionenforschung, 55128 Mainz, Germany}

\author{Felipe Pedreros Bustos}
\affiliation{Johannes Gutenberg-Universit{\"a}t Mainz, 55128 Mainz, Germany}
 \affiliation{Helmholtz-Institut Mainz, GSI Helmholtzzentrum f{\"u}r Schwerionenforschung, 55128 Mainz, Germany}
 \affiliation{Aix Marseille Univ, CNRS, CNES, LAM, Marseille, France}

\author{Alexander Akulshin}
\affiliation{Johannes Gutenberg-Universit{\"a}t Mainz, 55128 Mainz, Germany}
 \affiliation{Helmholtz-Institut Mainz, GSI Helmholtzzentrum f{\"u}r Schwerionenforschung, 55128 Mainz, Germany}
\affiliation{Optical Sciences Centre, Swinburne University of Technology, Melbourne 3122, Australia}
 
 \author{Hong Guo}
 \email{hongguo@pku.edu.cn}
\affiliation{State Key Laboratory of Advanced Optical Communication Systems and Networks, Department of Electronics, and Center for Quantum Information Technology, Peking University, Beijing 100871, China}

\author{Arne Wickenbrock}
\affiliation{Johannes Gutenberg-Universit{\"a}t Mainz, 55128 Mainz, Germany}
 \affiliation{Helmholtz-Institut Mainz, GSI Helmholtzzentrum f{\"u}r Schwerionenforschung, 55128 Mainz, Germany}

\author{Dmitry Budker}
\email{budker@uni-mainz.de}
\affiliation{Johannes Gutenberg-Universit{\"a}t Mainz, 55128 Mainz, Germany}
 \affiliation{Helmholtz-Institut Mainz, GSI Helmholtzzentrum f{\"u}r Schwerionenforschung, 55128 Mainz, Germany}
\affiliation{Department of Physics, University of California, Berkeley, California 94720, USA}

\date{\today}

\begin{abstract}

Stand-off magnetometry allows measuring magnetic field at a distance, and can be employed in geophysical research, hazardous environment monitoring, and security applications.
Stand-off magnetometry based on resonant scattering from atoms or molecules is often limited by the scarce amounts of detected light.
The situation would be dramatically improved if the light emitted by excited atoms were to propagate towards the excitation-light-source in a directional manner. Here, we demonstrate that this is possible by means of mirrorless lasing.
In a tabletop experiment, we detect free-precession signals of ground-state sodium spins under the influence of a magnetic field by measuring backward-directed light. This method enables scalar magnetometry in the Earth field range, that can be extended to long-baseline sensing.
\end{abstract}

\maketitle




Stand-off magnetometry, in which most of the experimental apparatus can be remote from the place where the magnetic field is measured, provides opportunity for remote geophysical research or hazardous environment monitoring~\cite{Fu2020Sensitive}.
Stand-off magnetometry based on resonant scattering from atoms, in particular, laser-guide-star (LGS)-based magnetometry~\cite{Higbie2011,Kane2018,Pedreros2018,Fan2019remote,Pedreros2018b}, where fluorescence-collection efficiency for a typical telescope is less than $10^{-10}$ or even lower, is often limited by the scarce number of detected photons. The problem could be solved if the excited atoms were forced to radiate towards the source in the form of laser-like emission. This is, indeed, possible using the technique of mirrorless lasing, a method to generate backward directional emission with the help of so called backward amplified spontaneous emission (ASE)~\cite{Thompson2014Pulsed,Akulshin2018}, which was demonstrated with atomic vapors confined to vapor cells. The question of the possibility to ``scale'' the laboratory results to on-sky experiments remains open and is a subject of ongoing investigation~\cite{Akulshin2018,hickson2020amplified}.

To apply mirrorless lasing to stand-off magnetometry, the central question is how to extract magnetic information from the backward emission.
It was previously shown that the threshold of mirrorless lasing and intensity of directional radiation are sensitive to the magnetic field in the interaction region~\cite{Akulshin2018}. However, since the intensity of directional radiation is also a complex nonlinear function of many experimental parameters such as laser-beam intensity, laser frequency, atomic vapor density, shape and dimensions of the excitation region, etc., determining the strength of magnetic fields from the intensity of directional light requires detailed calibration of this dependence that is practically impossible in the case of remote detection.

In this work, we suggested and implemented a technique of mirrorless-lasing-based stand-off magnetometry, which is absolute and calibration-free, and conducted successful proof-of-principle laboratory experiments with a sodium vapor cell.
  The method is based on the principle of "free" evolution of the ground state atomic spins under the influence of the magnetic field~\cite{Budker2007Optical}.
 To implement this, we perform the measurement in three stages: 1. Optical pumping of the ground-state spin polarization; 2. Free evolution of the atomic spins ``in the dark''; 3. Read out via mirrorless lasing, the intensity of which is related to the ground-state spin polarization.
 As the mirrorless lasing process requires high-intensity light to excite the atoms and generate population inversion between specific excited states, the ground-state spin polarization is strongly affected by this light. To reconstruct the full evolution of spin polarization, we repeat this measurement and combine the results with different evolution times. The resulting spin-precession signal oscillating at the Larmor frequency enables scalar magnetometry. Considering the fact that the returned emission does not require a retroreflector \cite{patton2012remotely}, our method is promising for adjustment-free, long-baseline remote magnetometry.

\begin{figure*}
\centering
\includegraphics[width=2\columnwidth]{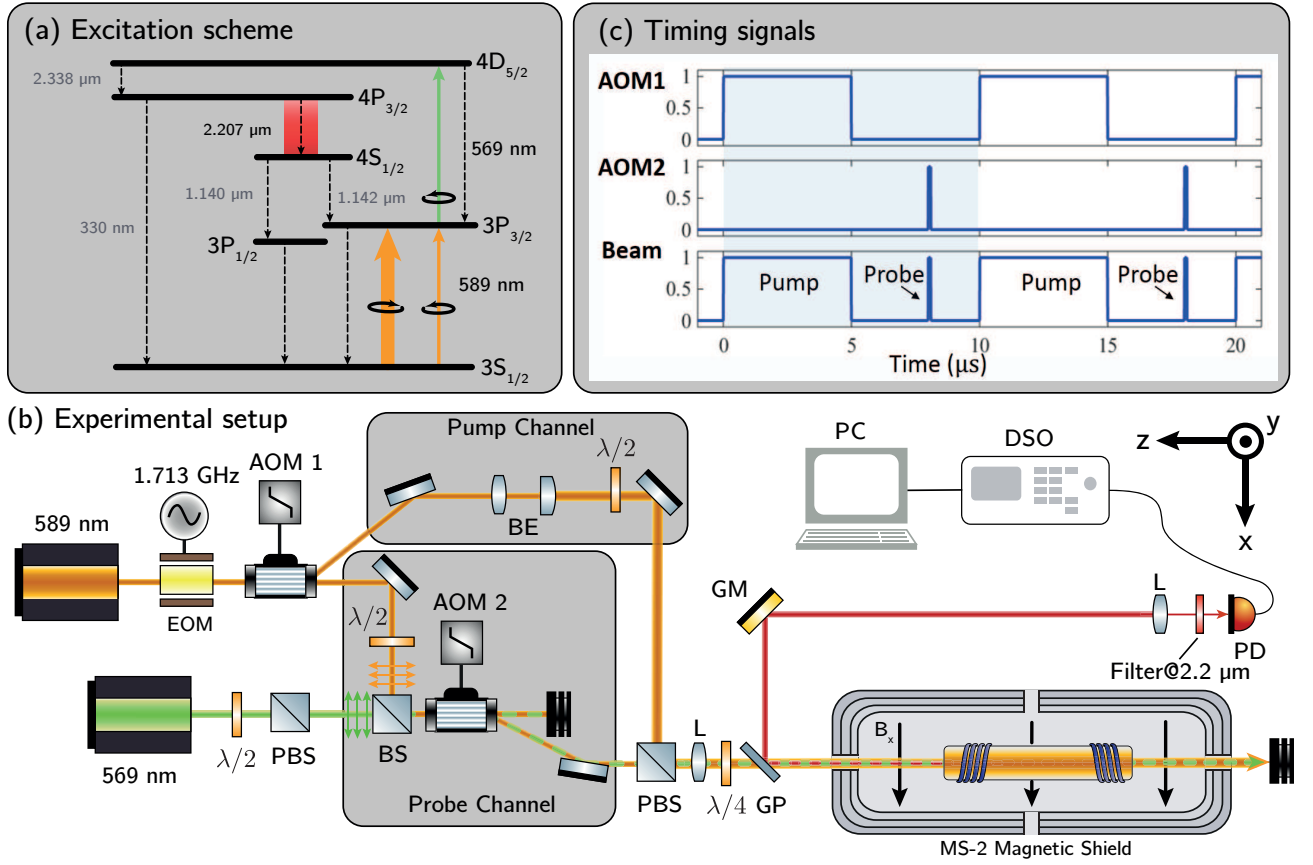}
    \caption{\textbf{Schematic of stand-off magnetometer.}
        \textbf{(a)} Energy levels of sodium atoms and excitation scheme. Pump beam is a left circularly polarized 589 nm laser exciting the 3$S_{1/2}$ to 3$P_{3/2}$ transition, which generates atomic spin polarization in the 3$S_{1/2}$ state; the probe beam is a combination of right circularly polarized 589 nm and 569 nm lights, which excites atoms to 4$D_{5/2}$ and therefore produces population inversion between 4$P_{3/2}$ and 4$S_{1/2}$, thus enabling gain for 2.21 $\mu$m spontaneous emission propagating backwards and generating a laser-like emission, or mirrorless lasing.
    \textbf{(b)} Experimental setup.  EOM, electro-optic modulator driven by 1.713 GHz signal to generate a sideband for hyperfine repumping; AOM1, acousto-optic modulator
used to pulse the pump beam; AOM2, acousto-optic modulator
used to pulse the probe beam; BE, beam expander; $\lambda/2$, half-wave plate; $\lambda/4$, quater-wave plate; PBS, polarizing beam splitter; BS, beam splitter; L, a convex lens; GP, glass plate for reflecting the mirrorless-lasing emission; GM, gold coated mirror; PD, photodiode; DSO, digital storage oscilloscope; PC, personal computer.
    \textbf{(c)} An example of the timing diagram. The figures at the top and the middle are the driving signals of AOM1 and AOM2 respectively, and the figure on the bottom represents the light pulses sent into the vapor cell.
 }\label{Fig_1}
\end{figure*}

The experimental arrangement is shown in Fig.~\ref{Fig_1}. Light from a Coherent CR-699 dye laser tuned to the sodium D$_2$-line ($3\text{S}_{1/2} \rightarrow 3\text{P}_{3/2}$ transition), see Fig.~\ref{Fig_1}(a), passes through an electro-optic modulator (EOM), see Fig.~\ref{Fig_1}(b), to produce sidebands used for hyperfine repumping. The beam then passes through an acousto-optic modulator (AOM1) that can switch the light power between the zeroth and first diffraction orders. The light from the first diffraction order is steered into a buffer-gas-free sodium vapor cell (3~cm diameter and 10~cm length) to generate atomic spin polarization in the ground state. The zeroth diffraction order beam is combined with the light from another dye laser (Coherent CR-899) tuned to 569 nm.
The combined beam is passed through a second AOM (AOM2) which is used a switch to control the optical excitation of the polarized ground state at a specific time. The probe beam excites ground-state atoms up to the $4\text{D}_{5/2}$ state enabling mirrorless lasing from population inverted transitions. Mirrorless-lasing emission at 2.21 $\mu$m is detected in the backward direction~\cite{Akulshin2018} with an InGaAs photodetector (Thorlabs PDA10D2, 25 MHz bandwidth) placed 1.5~m away from the center of the vapor cell.
The timing diagram is depicted in Fig.~\ref{Fig_1}(c). The driving signals to AOM1 and AOM2 are pulsed at the same frequency. Each period starts with a pulse of several $\mu$s duration sent to AOM1, which generates a pump-beam pulse to polarize the sodium atoms. After the end of the pump-beam pulse, the spin polarization evolves freely. After some time, a shorter pulse is sent to AOM2, generating a probe-beam pulse and a subsequent mirrorless-lasing emission. As the mirrorless lasing signal is a pulse train repeated at the modulation frequency, we accumulate the signal in a digital storage oscilloscope and then analyze the data to extract the pulse amplitude, which reflects the state of the spin polarization.

\begin{figure*}
\centering
\includegraphics[width=2\columnwidth]{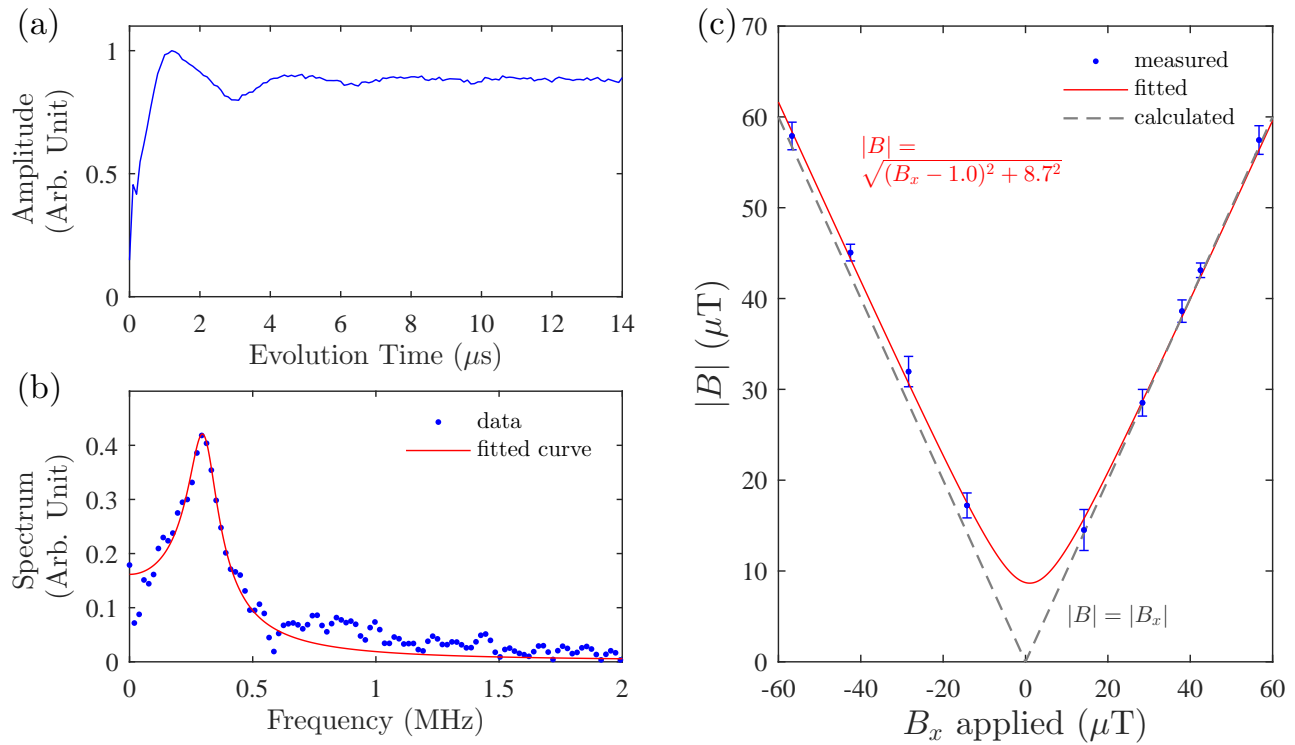}
    \caption{
    \textbf{(a)} Free precession signal with 42.5~${\mu}$T bias magnetic field applied along the $x$ direction. \textbf{(b)} Blue points: amplitude spectrum of the free precession signal shown in (a); red line: a Lorentz fitting. \textbf{(c)} Blue points: measured magnetic fields at different $B_x$ applied, which are derived from the central frequencies of the Lorentz fitting, as shown in (b); red line: a fitting of the measured magnetic fields, which suggests a 1.0(0.4)~$\mu$T residual magnetic field along -$x$ direction and a 8.7(1.2)~$\mu$T residual magnetic field in the $y-o-z$ plane in the magnetic shield; gray dashed line: a line that assumes no residual magnetic field in the magnetic shield, so that the total magnetic strength $|B|$ equals to the applied $|B_x|$.
 }\label{Fig_2}
\end{figure*}

The energy levels of sodium atoms and the excitation scheme are shown in Fig.~\ref{Fig_1}(a). The pump beam at 589 nm is resonant with the sodium D2 transition (from $2\text{S}_{1/2},\ F=2$  to $3\text{P}_{3/2},\  F'=1,2,3$). It is left circularly polarized such that it polarizes the sodium ground state spin. Probing with mirroless lasing is similar to that in Ref.~\cite{Akulshin2018}, in which the atoms in the $3\text{S}_{1/2}$ ground state were pumped into the $3\text{P}_{3/2}$ state with right circularly polarized 589~nm light and some of them were further pumped to the $4\text{D}_{5/2}$ state with right circularly polarized 569~nm light. Atoms in the $4\text{D}_{5/2}$ state decay into $4\text{P}_{3/2}$ which results in population inversion between $4\text{P}_{3/2}$ and $4\text{S}_{1/2}$. Such population inversion provides gain for 2.21
~$\mu$m spontaneous emission which propagates back to the photodetector. It is important for our technique that the intensity of the backward emission depends on the polarization of the ground state.

A typical spin-evolution signal detected with mirrorless lasing, Fig.~\ref{Fig_2}(a), shows an oscillating decay pattern. For this measurement, we have a modulation period of 20~$\mu$s, a pump pulse width of 5~$\mu$s and a probe pulse width of 0.4~$\mu$s. The peak power of the pump beam is 173~mW, and the peak power of the 589~nm and 569~nm components of the probe beam is 48~mW and 49~mW, respectively. The repumping sideband of 589 nm laser takes about 20$\%$ of the total power. The waist dimensions of the pump and probe beams at the center of the sodium cell are about 1.5$\times$3~mm$^2$ and 1$\times$1~mm$^2$, respectively. The vapor cell is heated to $\sim$169$^\circ$C. We scan the evolution time of the spin polarization from 0~$\mu$s to 14~$\mu$s with a 0.1~$\mu$s step.
The oscillating frequency of the evolution signal is directly determined by the Larmor frequency of sodium ground state and the decay rate is determined by the transit time of sodium atoms across the pump beam.
The signal starts from a lowest point. This is because the pump and probe beams are of opposite polarizations and the atoms are prepared in an unfavored state for mirrorless lasing during the pump stage. Then, as the spin polarization precess, the mirrorless lasing signal increases and reaches a peak when the polarization precesses for about 180$^\circ$.
After that, the oscillating amplitude decays as the optically pumped atoms escape from the pump region and the averaged polarization decreases. When the evolution time increases to about 12~$\mu$s, the signal reaches a steady state, corresponding to no remaining polarization before detection.
The amplitude spectrum of this evolution signal is shown in Fig.\,\ref{Fig_2}(b), in which there is a peak at 0.30\,MHz, corresponding to the Larmor frequency at 43~$\mu$T (the gyromagnetic ratio of $2\text{S}_{1/2}\ F=2$ level is $\approx$7\,Hz/nT). The amplitude spectrum deviates from the Lorentz fitting for frequency higher than 0.6 MHz, which may be due to the contributions from high-order harmonics, as there may be some nonlinear dependence of the mirrorless lasing signal on the atomic polarization.

To test the magnetic-field dependence of this peak frequency, we measure the evolution signal at different magnetic fields $B_x$ and use the peak frequency from the Lorentz fitting to estimate the magnetic field strength, with the results shown in Fig.\,\ref{Fig_2}(c). According to a fitting represented by the red line, there is a 1.0(0.4)~$\mu$T residual magnetic field along the -$x$ direction and a 8.7(1.2) ~$\mu$T residual magnetic field in the $y-o-z$ plane in the magnetic shield. Such magnetic-field dependence of this peak frequencies suggests that this method could be used for magnetometry.

In conclusion, we have demonstrated scalar stand-off magnetometry based on 2.21~$\mu$m mirrorless lasing with a buffer-gas-free sodium vapor cell. By operating in three stages as optical pumping, dark evolution and mirrorless lasing detection, we observed free-precession signals of sodium ground-state polarization. The complex signal may have some vector information in it, which can be investigated in the future. Compared with conventional fluorescence detection, the detection of backward directional emission in this work is expected to dramatically improve the efficiency of light collection and thus the magnetometry sensitivity.  Vapor cells filled with buffer gas such as He or Ne are expected to exhibit much longer coherence times as the atoms would spend more time in the light-beam area, which is beneficial to high-sensitivity magnetometry.  Our method can potentially also be extended to other mirrorless lasing schemes or other atoms. Ongoing work is exploring the feasibility of such schemes for remote geophysical magnetometry~\cite{Pedreros2018,Pedreros2018b}.

\section*{Acknowledgement}
The authors are grateful to Aram Papoyan for helpful discussions, as well as to the European Southern Observatory for the loan of a dye-laser system. This work was supported in part by the DFG Project ID 390831469:  EXC 2118 (PRISMA+ Cluster of Excellence). DB also received support from the European Research Council (ERC) under the European Union Horizon 2020 Research and Innovation Program (grant agreement No. 695405) and from the DFG Reinhart Koselleck Project.  RZ acknowledges support from the China
Scholarship Council (CSC) enabling his research at the
Helmholtz-Institut Mainz.
FPB received support from the European Union's Horizon 2020 research and innovation programme under the Marie Sklodowska-Curie grant agreement No 893150.

\bibliography{Main Text_V1.1.bib}

\begin{thebibliography}{11}%
\makeatletter
\providecommand \@ifxundefined [1]{%
 \@ifx{#1\undefined}
}%
\providecommand \@ifnum [1]{%
 \ifnum #1\expandafter \@firstoftwo
 \else \expandafter \@secondoftwo
 \fi
}%
\providecommand \@ifx [1]{%
 \ifx #1\expandafter \@firstoftwo
 \else \expandafter \@secondoftwo
 \fi
}%
\providecommand \natexlab [1]{#1}%
\providecommand \enquote  [1]{``#1''}%
\providecommand \bibnamefont  [1]{#1}%
\providecommand \bibfnamefont [1]{#1}%
\providecommand \citenamefont [1]{#1}%
\providecommand \href@noop [0]{\@secondoftwo}%
\providecommand \href [0]{\begingroup \@sanitize@url \@href}%
\providecommand \@href[1]{\@@startlink{#1}\@@href}%
\providecommand \@@href[1]{\endgroup#1\@@endlink}%
\providecommand \@sanitize@url [0]{\catcode `\\12\catcode `\$12\catcode
  `\&12\catcode `\#12\catcode `\^12\catcode `\_12\catcode `\%12\relax}%
\providecommand \@@startlink[1]{}%
\providecommand \@@endlink[0]{}%
\providecommand \url  [0]{\begingroup\@sanitize@url \@url }%
\providecommand \@url [1]{\endgroup\@href {#1}{\urlprefix }}%
\providecommand \urlprefix  [0]{URL }%
\providecommand \Eprint [0]{\href }%
\providecommand \doibase [0]{https://doi.org/}%
\providecommand \selectlanguage [0]{\@gobble}%
\providecommand \bibinfo  [0]{\@secondoftwo}%
\providecommand \bibfield  [0]{\@secondoftwo}%
\providecommand \translation [1]{[#1]}%
\providecommand \BibitemOpen [0]{}%
\providecommand \bibitemStop [0]{}%
\providecommand \bibitemNoStop [0]{.\EOS\space}%
\providecommand \EOS [0]{\spacefactor3000\relax}%
\providecommand \BibitemShut  [1]{\csname bibitem#1\endcsname}%
\let\auto@bib@innerbib\@empty
\bibitem [{\citenamefont {Fu}\ \emph {et~al.}(2020)\citenamefont {Fu},
  \citenamefont {Iwata}, \citenamefont {Wickenbrock},\ and\ \citenamefont
  {Budker}}]{Fu2020Sensitive}%
  \BibitemOpen
  \bibfield  {author} {\bibinfo {author} {\bibfnamefont {K.-M.~C.}\
  \bibnamefont {Fu}}, \bibinfo {author} {\bibfnamefont {G.~Z.}\ \bibnamefont
  {Iwata}}, \bibinfo {author} {\bibfnamefont {A.}~\bibnamefont {Wickenbrock}},\
  and\ \bibinfo {author} {\bibfnamefont {D.}~\bibnamefont {Budker}},\
  }\bibfield  {title} {\bibinfo {title} {Sensitive magnetometry in challenging
  environments},\ }\href@noop {} {\bibfield  {journal} {\bibinfo  {journal}
  {AVS Quantum Sci.}\ }\textbf {\bibinfo {volume} {2}},\ \bibinfo {pages}
  {044702} (\bibinfo {year} {2020})}\BibitemShut {NoStop}%
\bibitem [{\citenamefont {Higbie}\ \emph {et~al.}(2011)\citenamefont {Higbie},
  \citenamefont {Rochester}, \citenamefont {Patton}, \citenamefont
  {Holzl{\"o}hner}, \citenamefont {Bonaccini~Calia},\ and\ \citenamefont
  {Budker}}]{Higbie2011}%
  \BibitemOpen
  \bibfield  {author} {\bibinfo {author} {\bibfnamefont {J.~M.}\ \bibnamefont
  {Higbie}}, \bibinfo {author} {\bibfnamefont {S.~M.}\ \bibnamefont
  {Rochester}}, \bibinfo {author} {\bibfnamefont {B.}~\bibnamefont {Patton}},
  \bibinfo {author} {\bibfnamefont {R.}~\bibnamefont {Holzl{\"o}hner}},
  \bibinfo {author} {\bibfnamefont {D.}~\bibnamefont {Bonaccini~Calia}},\ and\
  \bibinfo {author} {\bibfnamefont {D.}~\bibnamefont {Budker}},\ }\bibfield
  {title} {\bibinfo {title} {Magnetometry with mesospheric sodium},\ }\href
  {https://doi.org/10.1073/pnas.1013641108} {\bibfield  {journal} {\bibinfo
  {journal} {Proc. Natl. Acad. Sci.}\ }\textbf {\bibinfo {volume} {108}},\
  \bibinfo {pages} {3522} (\bibinfo {year} {2011})}\BibitemShut {NoStop}%
\bibitem [{\citenamefont {Kane}\ \emph {et~al.}(2018)\citenamefont {Kane},
  \citenamefont {Hillman}, \citenamefont {Denman}, \citenamefont {Hart},
  \citenamefont {Scott}, \citenamefont {Purucker},\ and\ \citenamefont
  {Potashnik}}]{Kane2018}%
  \BibitemOpen
  \bibfield  {author} {\bibinfo {author} {\bibfnamefont {T.~J.}\ \bibnamefont
  {Kane}}, \bibinfo {author} {\bibfnamefont {P.~D.}\ \bibnamefont {Hillman}},
  \bibinfo {author} {\bibfnamefont {C.~A.}\ \bibnamefont {Denman}}, \bibinfo
  {author} {\bibfnamefont {M.}~\bibnamefont {Hart}}, \bibinfo {author}
  {\bibfnamefont {R.~P.}\ \bibnamefont {Scott}}, \bibinfo {author}
  {\bibfnamefont {M.~E.}\ \bibnamefont {Purucker}},\ and\ \bibinfo {author}
  {\bibfnamefont {S.~J.}\ \bibnamefont {Potashnik}},\ }\bibfield  {title}
  {\bibinfo {title} {Laser remote magnetometry using mesospheric sodium},\
  }\href {https://doi.org/10.1029/2018JA025178} {\bibfield  {journal} {\bibinfo
   {journal} {J. Geophys. Res.: Space Physics}\ }\textbf {\bibinfo {volume}
  {123}},\ \bibinfo {pages} {6171} (\bibinfo {year} {2018})}\BibitemShut
  {NoStop}%
\bibitem [{\citenamefont {Pedreros~Bustos}\ \emph {et~al.}(2018)\citenamefont
  {Pedreros~Bustos}, \citenamefont {Bonaccini~Calia}, \citenamefont {Budker},
  \citenamefont {Centrone}, \citenamefont {Hellemeier}, \citenamefont
  {Hickson}, \citenamefont {Holzl{\"o}hner},\ and\ \citenamefont
  {Rochester}}]{Pedreros2018}%
  \BibitemOpen
  \bibfield  {author} {\bibinfo {author} {\bibfnamefont {F.}~\bibnamefont
  {Pedreros~Bustos}}, \bibinfo {author} {\bibfnamefont {D.}~\bibnamefont
  {Bonaccini~Calia}}, \bibinfo {author} {\bibfnamefont {D.}~\bibnamefont
  {Budker}}, \bibinfo {author} {\bibfnamefont {M.}~\bibnamefont {Centrone}},
  \bibinfo {author} {\bibfnamefont {J.}~\bibnamefont {Hellemeier}}, \bibinfo
  {author} {\bibfnamefont {P.}~\bibnamefont {Hickson}}, \bibinfo {author}
  {\bibfnamefont {R.}~\bibnamefont {Holzl{\"o}hner}},\ and\ \bibinfo {author}
  {\bibfnamefont {S.}~\bibnamefont {Rochester}},\ }\bibfield  {title} {\bibinfo
  {title} {Remote sensing of geomagnetic fields and atomic collisions in the
  mesosphere},\ }\href {https://doi.org/10.1038/s41467-018-06396-7} {\bibfield
  {journal} {\bibinfo  {journal} {Nature Communications}\ }\textbf {\bibinfo
  {volume} {9}},\ \bibinfo {pages} {3981} (\bibinfo {year} {2018})}\BibitemShut
  {NoStop}%
\bibitem [{\citenamefont {Fan}\ \emph {et~al.}(2019)\citenamefont {Fan},
  \citenamefont {Yang}, \citenamefont {Dong}, \citenamefont {Zhang},
  \citenamefont {Cui}, \citenamefont {Qian}, \citenamefont {Dong},
  \citenamefont {Deng}, \citenamefont {Zhou}, \citenamefont {Wei},
  \citenamefont {Feng},\ and\ \citenamefont {Chen}}]{Fan2019remote}%
  \BibitemOpen
  \bibfield  {author} {\bibinfo {author} {\bibfnamefont {T.}~\bibnamefont
  {Fan}}, \bibinfo {author} {\bibfnamefont {X.}~\bibnamefont {Yang}}, \bibinfo
  {author} {\bibfnamefont {J.}~\bibnamefont {Dong}}, \bibinfo {author}
  {\bibfnamefont {L.}~\bibnamefont {Zhang}}, \bibinfo {author} {\bibfnamefont
  {S.}~\bibnamefont {Cui}}, \bibinfo {author} {\bibfnamefont {J.}~\bibnamefont
  {Qian}}, \bibinfo {author} {\bibfnamefont {R.}~\bibnamefont {Dong}}, \bibinfo
  {author} {\bibfnamefont {K.}~\bibnamefont {Deng}}, \bibinfo {author}
  {\bibfnamefont {T.}~\bibnamefont {Zhou}}, \bibinfo {author} {\bibfnamefont
  {K.}~\bibnamefont {Wei}}, \bibinfo {author} {\bibfnamefont {Y.}~\bibnamefont
  {Feng}},\ and\ \bibinfo {author} {\bibfnamefont {W.}~\bibnamefont {Chen}},\
  }\bibfield  {title} {\bibinfo {title} {Remote magnetometry with mesospheric
  sodium based on gated photon counting},\ }\href
  {https://doi.org/https://doi.org/10.1029/2019JA026956} {\bibfield  {journal}
  {\bibinfo  {journal} {J. Geophys. Res.: Space Physics}\ }\textbf {\bibinfo
  {volume} {124}},\ \bibinfo {pages} {7505} (\bibinfo {year}
  {2019})}\BibitemShut {NoStop}%
\bibitem [{\citenamefont {{Pedreros Bustos}}\ \emph {et~al.}(2018)\citenamefont
  {{Pedreros Bustos}}, \citenamefont {{Bonaccini Calia}}, \citenamefont
  {{Budker}}, \citenamefont {{Centrone}}, \citenamefont {{Hellemeier}},
  \citenamefont {{Hickson}}, \citenamefont {{Holzl{\"o}hner}},\ and\
  \citenamefont {{Rochester}}}]{Pedreros2018b}%
  \BibitemOpen
  \bibfield  {author} {\bibinfo {author} {\bibfnamefont {F.}~\bibnamefont
  {{Pedreros Bustos}}}, \bibinfo {author} {\bibfnamefont {D.}~\bibnamefont
  {{Bonaccini Calia}}}, \bibinfo {author} {\bibfnamefont {D.}~\bibnamefont
  {{Budker}}}, \bibinfo {author} {\bibfnamefont {M.}~\bibnamefont
  {{Centrone}}}, \bibinfo {author} {\bibfnamefont {J.}~\bibnamefont
  {{Hellemeier}}}, \bibinfo {author} {\bibfnamefont {P.}~\bibnamefont
  {{Hickson}}}, \bibinfo {author} {\bibfnamefont {R.}~\bibnamefont
  {{Holzl{\"o}hner}}},\ and\ \bibinfo {author} {\bibfnamefont {S.}~\bibnamefont
  {{Rochester}}},\ }\bibfield  {title} {\bibinfo {title} {{Polarization-driven
  spin precession of mesospheric sodium atoms}},\ }\href
  {https://doi.org/10.1364/OL.43.005825} {\bibfield  {journal} {\bibinfo
  {journal} {Opt. Lett.}\ }\textbf {\bibinfo {volume} {43}},\ \bibinfo {pages}
  {5825} (\bibinfo {year} {2018})}\BibitemShut {NoStop}%
\bibitem [{\citenamefont {Thompson}\ \emph {et~al.}(2014)\citenamefont
  {Thompson}, \citenamefont {Ballmann}, \citenamefont {Cai}, \citenamefont
  {Yi}, \citenamefont {Rostovtsev}, \citenamefont {Sokolov}, \citenamefont
  {Hemmer}, \citenamefont {Zheltikov}, \citenamefont {Ariunbold},\ and\
  \citenamefont {Scully}}]{Thompson2014Pulsed}%
  \BibitemOpen
  \bibfield  {author} {\bibinfo {author} {\bibfnamefont {J.~V.}\ \bibnamefont
  {Thompson}}, \bibinfo {author} {\bibfnamefont {C.~W.}\ \bibnamefont
  {Ballmann}}, \bibinfo {author} {\bibfnamefont {H.}~\bibnamefont {Cai}},
  \bibinfo {author} {\bibfnamefont {Z.}~\bibnamefont {Yi}}, \bibinfo {author}
  {\bibfnamefont {Y.~V.}\ \bibnamefont {Rostovtsev}}, \bibinfo {author}
  {\bibfnamefont {A.~V.}\ \bibnamefont {Sokolov}}, \bibinfo {author}
  {\bibfnamefont {P.}~\bibnamefont {Hemmer}}, \bibinfo {author} {\bibfnamefont
  {A.~M.}\ \bibnamefont {Zheltikov}}, \bibinfo {author} {\bibfnamefont {G.~O.}\
  \bibnamefont {Ariunbold}},\ and\ \bibinfo {author} {\bibfnamefont {M.~O.}\
  \bibnamefont {Scully}},\ }\bibfield  {title} {\bibinfo {title} {Pulsed
  cooperative backward emissions from non-degenerate atomic transitions in
  sodium},\ }\href {https://doi.org/10.1088/1367-2630/16/10/103017} {\bibfield
  {journal} {\bibinfo  {journal} {New Journal of Physics}\ }\textbf {\bibinfo
  {volume} {16}},\ \bibinfo {pages} {103017} (\bibinfo {year}
  {2014})}\BibitemShut {NoStop}%
\bibitem [{\citenamefont {Akulshin}\ \emph {et~al.}(2018)\citenamefont
  {Akulshin}, \citenamefont {{Pedreros Bustos}},\ and\ \citenamefont
  {Budker}}]{Akulshin2018}%
  \BibitemOpen
  \bibfield  {author} {\bibinfo {author} {\bibfnamefont {A.~M.}\ \bibnamefont
  {Akulshin}}, \bibinfo {author} {\bibfnamefont {F.}~\bibnamefont {{Pedreros
  Bustos}}},\ and\ \bibinfo {author} {\bibfnamefont {D.}~\bibnamefont
  {Budker}},\ }\bibfield  {title} {\bibinfo {title} {Continuous-wave mirrorless
  lasing at 2.21 $\mu$m in sodium vapors},\ }\href
  {https://doi.org/10.1364/OL.43.005279} {\bibfield  {journal} {\bibinfo
  {journal} {Opt. Lett.}\ }\textbf {\bibinfo {volume} {43}},\ \bibinfo {pages}
  {5279} (\bibinfo {year} {2018})}\BibitemShut {NoStop}%
\bibitem [{\citenamefont {Hickson}\ \emph {et~al.}(2020)\citenamefont
  {Hickson}, \citenamefont {Hellemeier},\ and\ \citenamefont
  {Yang}}]{hickson2020amplified}%
  \BibitemOpen
  \bibfield  {author} {\bibinfo {author} {\bibfnamefont {P.}~\bibnamefont
  {Hickson}}, \bibinfo {author} {\bibfnamefont {J.}~\bibnamefont
  {Hellemeier}},\ and\ \bibinfo {author} {\bibfnamefont {R.}~\bibnamefont
  {Yang}},\ }\href@noop {} {\bibinfo {title} {Can amplified spontaneous
  emission produce intense laser guide stars for adaptive optics?}} (\bibinfo
  {year} {2020}),\ \Eprint {https://arxiv.org/abs/2012.11946} {arXiv:2012.11946
  [astro-ph.IM]} \BibitemShut {NoStop}%
\bibitem [{\citenamefont {Budker}\ and\ \citenamefont
  {Romalis}(2007)}]{Budker2007Optical}%
  \BibitemOpen
  \bibfield  {author} {\bibinfo {author} {\bibfnamefont {D.}~\bibnamefont
  {Budker}}\ and\ \bibinfo {author} {\bibfnamefont {M.}~\bibnamefont
  {Romalis}},\ }\bibfield  {title} {\bibinfo {title} {Optical magnetometry},\
  }\href {https://doi.org/10.1038/nphys566} {\bibfield  {journal} {\bibinfo
  {journal} {Nat. Phys.}\ }\textbf {\bibinfo {volume} {3}},\ \bibinfo {pages}
  {227} (\bibinfo {year} {2007})}\BibitemShut {NoStop}%
\bibitem [{\citenamefont {Patton}\ \emph {et~al.}(2012)\citenamefont {Patton},
  \citenamefont {Versolato}, \citenamefont {Hovde}, \citenamefont {Corsini},
  \citenamefont {Higbie},\ and\ \citenamefont {Budker}}]{patton2012remotely}%
  \BibitemOpen
  \bibfield  {author} {\bibinfo {author} {\bibfnamefont {B.}~\bibnamefont
  {Patton}}, \bibinfo {author} {\bibfnamefont {O.}~\bibnamefont {Versolato}},
  \bibinfo {author} {\bibfnamefont {D.~C.}\ \bibnamefont {Hovde}}, \bibinfo
  {author} {\bibfnamefont {E.}~\bibnamefont {Corsini}}, \bibinfo {author}
  {\bibfnamefont {J.~M.}\ \bibnamefont {Higbie}},\ and\ \bibinfo {author}
  {\bibfnamefont {D.}~\bibnamefont {Budker}},\ }\bibfield  {title} {\bibinfo
  {title} {A remotely interrogated all-optical {$^{87}$Rb} magnetometer},\
  }\href@noop {} {\bibfield  {journal} {\bibinfo  {journal} {Applied Physics
  Letters}\ }\textbf {\bibinfo {volume} {101}},\ \bibinfo {pages} {083502}
  (\bibinfo {year} {2012})}\BibitemShut {NoStop}%
\end{thebibliography}%

\end{document}